# Ferromagnetic Enhancement in LaMnO$_3$ Films with Release and Flexure

Hongbao Yao[1,2], Kuijuan Jin[1,2,3,*], Zhen Yang[1,2], Qinghua Zhang[1], Wenning Ren[1,2], Shuai Xu[1,2], Mingwei Yang[1,2], Lin Gu[1,2], Er-Jia Guo[1,2,3], Chen Ge[1,2], Can Wang[1,2,3], Xiulai Xu[1,2,3], Dongxiang Zhang[1,2], and Guozhen Yang[1,2]

[1] Beijing National Laboratory for Condensed Matter Physics, Institute of Physics, Chinese Academy of Sciences, Beijing 100190, China
[2] University of Chinese Academy of Sciences, Beijing 100049, China
[3] Songshan Lake Materials Laboratory, Dongguan, Guangdong 523808, China

*Author to whom correspondence should be addressed: kjjin@iphy.ac.cn

**Abstract**

A variety of novel phenomena and functionalities emerge from lowering the dimensionality of materials and enriching the degrees of freedom in modulation. In this work, it is found that the saturation magnetization of LaMnO$_3$ (LMO) films is largely enhanced by 56% after releasing from a brand-new phase of tetragonal strontium aluminate buffer layer, and is significantly increased by 92% with bending films to a curvature of 1 mm$^{-1}$ using a water-assisted direct-transferring method. Meanwhile, the Curie temperature of LMO films has been improved by 13 K. High-resolution spherical aberration-corrected scanning transmission electron microscopy and first-principles calculations unambiguously demonstrate that the enhanced ferromagnetism is attributed to the strengthened Mn-O-Mn super-exchange interactions from the augmented characteristics of the unconventional P2$_1$/n structure caused by the out-of-plane lattice shrinking after strain releasing and increased flexure degree of freestanding LMO films. This work paves a way to achieve large-scale and crack-and-wrinkle-free freestanding films of oxides with largely improved functionalities.

**[Keywords]** Freestanding manganite, P2$_1$/n LaMnO$_3$, flexomagnetism

## 1. Introduction

Substrate-free two-dimensional (2D) materials have ignited broad interests for the observation of unprecedented phases and properties emerging from manipulating the coupling among the degrees of charge, lattice, orbital, and spin by mechanical regulations, such as bending, rotating, and twisting.[1-3] If applied to 2D perovskites, these mechanical regulations make perovskite materials versatile testbed to discover new phenomena, mechanisms, and functionalities.[4,5] Recently, the application of water-soluble substrates (NaCl, KCl, etc.) and buffer layers ($Sr_3Al_2O_6$, BaO, etc.) in perovskite heteroepitaxy has enabled the accomplishment of artificial 2D freestanding perovskite films and has granted researchers the ability to modulate their geometries and functionalities through extra means aside from conventional strain from substrates.[6-11] Although numerous exotic and distinctive phenomena and functionalities, such as super-elasticity,[7,8] boosted piezoelectricity,[9] and emerging ferroelectricity[10] in 2D freestanding perovskites have been discovered, the enhancement in the ferromagnetism of freestanding oxide materials and the acquisition of large-scale and crack-and-wrinkle-free freestanding perovskite films have rarely been reported.

On the other hand, the coexistence of ferromagnetism and insulating behavior in $LaMnO_3$ (LMO) has been well theoretically explained by Hou and Gong *et al*. through proposing an unconventional $P2_1/n$ structure,[12] which has been experimentally evidenced in our previous work.[13] However, the ferromagnetism in LMO is usually found to be restrained with out-of-plane lattice shrinking for films on the substrates.[14-18] Here we report a novel and significant enhancement of the ferromagnetism of LMO films with releasing and bending, while the out-of-plane lattice was shrunk. High-resolution spherical aberration-corrected scanning transmission electron microscopy (STEM) and the first-principles calculations demonstrated the enhancement of the ferromagnetism was attributed to the strengthening of the Mn-O-Mn super-exchange interactions caused by the more distinguished characteristics of the $P2_1/n$ structure with out-of-plane lattice shrinking after releasing and bending. This out-of-plane lattice shrinking in above processes agrees well with the indications from the X-ray diffraction and Raman spectra.

**2. Results and Discussion**

The multilayer heterostructure was fabricated by a successive growth of a brand-new phase of tetragonal strontium aluminate (SAO) layer (Figure S1a,b, Supporting Information), rather than the cubic one by other groups,[6-8,19-22] on the TiO$_2$-termintated (001) SrTiO$_3$ (STO) substrate, followed by an *in-situ* LMO layer (see Experimental Section). The phase purity and excellent crystallinity of the subsequent LMO layer were unveiled by the X-ray diffraction (XRD) (**Figure 1**a) and its rocking curve (Inset in Figure 1a), respectively, which agreed well with the coherent growth feature from X-ray reciprocal space mapping (RSM) (Figure 1b). To realize the complete transferring of the LMO layer, we elaborately developed an auxiliary-free transferring method (Figure S2, Supporting Information), with which we were able to obtain massive large-scale and crack-and-wrinkle-free freestanding films and heterostructures and to transfer them onto arbitrary base at arbitrary angle. The floating LMO layers were transferred onto diamagnetic quartz rods with a curvature of $K = 0$ mm$^{-1}$, 0.5 mm$^{-1}$, and 1 mm$^{-1}$, respectively, to induce different degrees of flexure. A schematic illustration of the transferred (curved) LMO layer is shown in Figure 1c. The LMO only showed the (00*l*) peaks (Figure 1a), and maintained a pure phase and crystallinity after transferring, indicated by the RSM of the freestanding LMO (FSLMO) with $K = 0$ mm$^{-1}$ (Figure 1d), and exhibited continuous decrease in lattice *c* (Figure 1e), which implied the LMO membranes were subjected to monotonous out-of-plane lattice shrinking during above processes. This shrinking was consistent with the increase of the in-plane lattice parameters of LMO after the transferring, evaluated from Figure S3a,b, Supporting Information, and listed in Table S1, Supporting Information. The thickness and roughness of LMO, revealed by X-ray reflectivity (XRR), were 100 u.c. (Figure 1f) and 5.7 Å (Figure S3c, Supporting Information), respectively, before transferring, and were 100 u.c. (Figure 1g) and 4.9 Å (Figure S3d, Supporting Information), respectively, after transferring. The identical thickness and roughness before and after transferring verified that there was approximately no loss in the quantity of LMO unit cells along *c* direction and in the surface topography during the transferring.

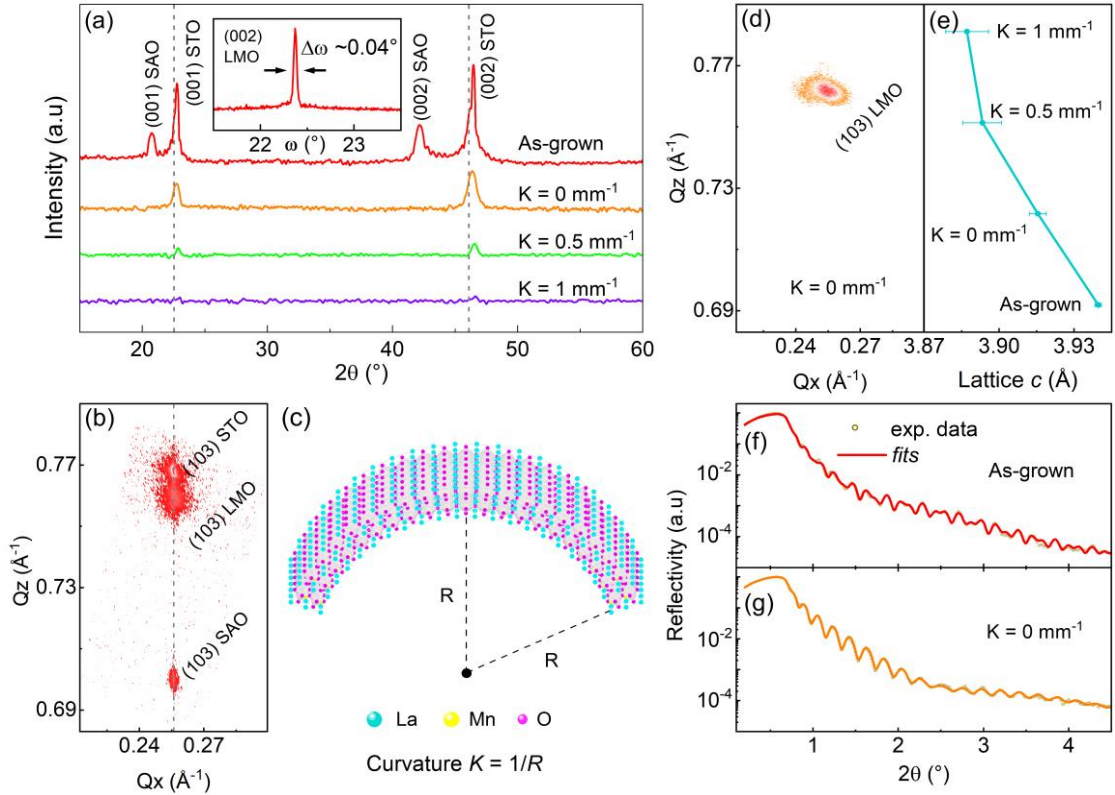

**Figure 1**. a) XRD of as-grown LMO and transferred FSLMO. The left and right dotted line denote the position of the (001) and (002) peak of as-grown LMO, respectively. Being not flat, the X-ray beam cannot be well-aligned to focus on the film, which may explain the drop of peak intensity for bended samples. Inset illustrates the rocking curve on the (002) peak of as-grown LMO. b) RSM of as-grown LMO. The dotted line denotes the position of the Qx of STO substrate. c) Schematic of a bended FSLMO where $R$ and $K$ represent the radius and curvature of the FSLMO, respectively. d) RSM of FSLMO with $K = 0$ mm$^{-1}$. e) Lattice parameter $c$ of LMO in (a). f, g) XRR of as-grown LMO and FSLMO with $K = 0$ mm$^{-1}$, respectively. The dots represent experimental data and the curves denote the best fittings.

**Figure 2** shows the evolution of the magnetic properties of LMO films after releasing and bending. The magnetic hysteresis (MH) loops of the as-grown LMO and bended FSLMO membranes, measured with the magnetic field applied parallel to $ab$ plane at 10 K, are presented in Figure 2a, which clearly shows the remarkable enhancement in saturation magnetization ($M$s) after releasing and bending for LMO films. Figure 2b shows the magnetization ($M$) dependence on temperature (MT) after a field cooling (FC) process under an applied magnetic field of 1000 Oe, and presents an accompanying appreciable increase in $T_c$, which was determined by best fitting to the paramagnetic region of the MT curves with Curie-Weiss law (Figure S4, Supporting Information). Such enhancement in $M_s$ and $T_c$ after releasing had also been observed in LMO films of different thicknesses (25 u.c. in Figure S5a,b, Supporting Information, and 50 u.c. in Figure S5c,d, Supporting Information). In detail, the $M_s$ of the LMO membrane boosted from 1.04 to 1.63 $\mu B \cdot Mn^{-1}$ after releasing, increased by an amplitude of 56%. Moreover, through varying the flexure degree in FSLMO, the $M_s$ further augmented to 1.8 ($K = 0.5$ mm$^{-1}$) and 2 $\mu B \cdot Mn^{-1}$ ($K = 1$ mm$^{-1}$), enhanced by a percentage of 92% comparing to that of as-grown LMO, as summarized in Figure 2c. Simultaneously, the $T_c$ increased from 123 to 130 K after releasing and further rase up to 133 ($K = 0.5$ mm$^{-1}$) and 136 K ($K = 1$ mm$^{-1}$), respectively, during the bending processes.

  Since the valent state of Mn plays a significant role in manganite magnetism,[13,15,23-25] we collected the unit-cell-resolved electron energy loss spectroscopy (EELS) spectra on Mn $L$ edge within 10 u.c. region from the bottom of LMO layer of the as-grown LMO, the FSLMO with $K = 0$ mm$^{-1}$, and that with $K = 1$ mm$^{-1}$, as displayed in Figure S6a,b, Supporting Information, and Figure 2d, respectively. The spectra of these samples present steady characteristic peaks of $Mn^{3+}$ at 641.7 ($L_3$) and 653.1 eV ($L_2$), and no sign of $Mn^{2+}$ and $Mn^{4+}$ can be distinguished. Moreover, the area ratios of $L_3/L_2$ for Mn in each atomic layer of the as-grown LMO and FSLMO, shown in Figure 2e, remain closely to 3. These results demonstrate that Mn was kept in +3 valent state in all samples[13,23,24] without valence change, which contributes to the enhancement of the ferromagnetism in LMO, in the releasing and bending processes. On the other hand, the enhancement of the ferromagnetism in LMO with out-of-plane lattice shrinking from releasing and bending, which is revealed by prior XRD analyses (Figure 1e) and is in accordance with the softening of out-of-phase stretching ($B_{3g}$) mode in Raman spectra (Figure 2f)[26-28], is quite intriguing, comparing to the ferromagnetic suppression with the lattice shrinking reported in many previous studies.[12-18,29,30]

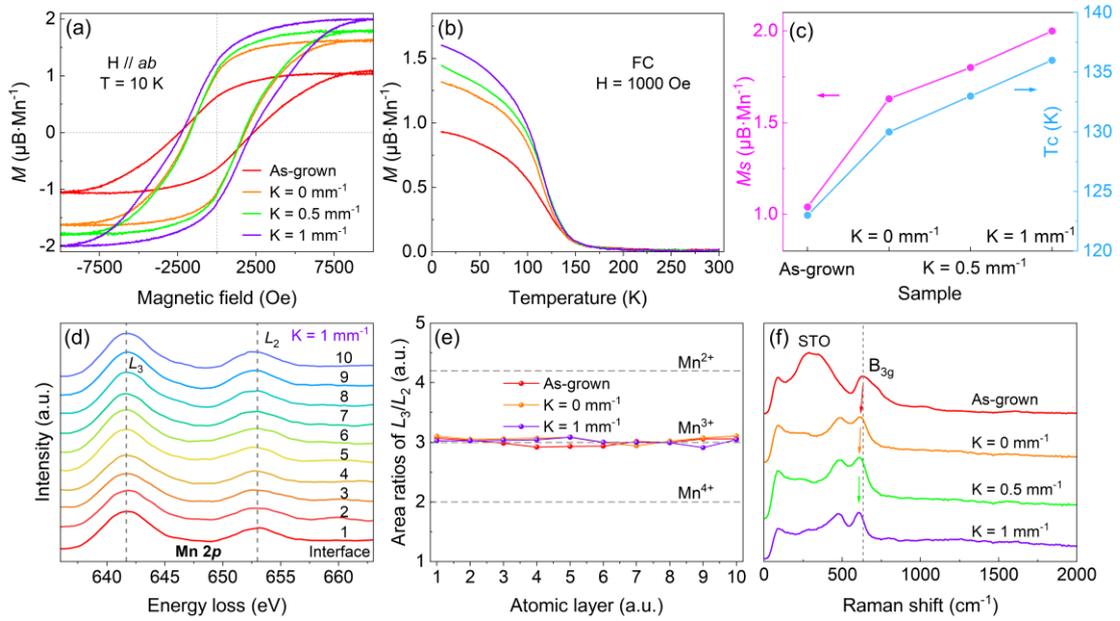

**Figure 2.** a) MH loops of as-grown LMO and FSLMO. b) Temperature-dependent magnetization corresponding to (a). c) Summary of $M_s$ and $T_c$ derivate from (a) and (b). d) EELS spectra collected over Mn $L$ edge of FSLMO with $K$ = 1 mm$^{-1}$. The dotted lines indicate the peak position of $L_3$ and $L_2$ edge of Mn$^{3+}$, respectively. Data has been normalized and offset in y-axis for better view. e) Evaluated $L_3/L_2$ area ratios of Mn in as-grown LMO, FSLMO with $K$ = 0 mm$^{-1}$, and that with $K$ = 1 mm$^{-1}$. The dashed lines represent the $L_3/L_2$ area ratios for pure Mn$^{2+}$, Mn$^{3+}$, and Mn$^{4+}$.[23,31] f) Raman shift for as-grown LMO and FSLMO. The dashed line represents the peak position of the B$_{3g}$ mode of the as-grown LMO.

To identify how the magnetic properties depend on the LMO structure microscopically after releasing and bending, the atomic structure of the as-grown LMO, the FSLMO with $K$ = 0 mm$^{-1}$, and that with $K$ = 1 mm$^{-1}$ were characterized with high-resolution spherical aberration-corrected STEM. **Figure 3**a shows the overview of the cross section of the as-grown LMO under high-angle annular dark-field (HAADF) mode and exhibits evident scarification of LMO, SAO, and STO with distinct contrasts. Figure 3b-f presents the elemental distribution of Sr, Al, La, Mn, and O, respectively, from EELS mapping corresponding to Figure 3a, which also exhibits clear elemental boundaries and indicates no transmembrane intermixing, interdiffusion, and chemical reactions in our samples.[19] Figure 3h displays the magnification of the marked area in Figure 3a and presents uniform atomic distributions and abrupt LMO-SAO interface under HAADF-STEM, verifying the coherent growth of LMO. The overview of the cross section of FSLMO with $K = 0$ mm$^{-1}$ is shown in Figure 3g and that of the FSLMO with $K = 1$ mm$^{-1}$ is presented in Figure S7a, Supporting Information, where the bended state of the LMO layer is distinguishable.

Figure 3i (Figure 3j and Figure S7b, Supporting Information) shows the high-resolution inverted annular bright-field (ABF) cross section at the bottom (bottom and top) of LMO layer of as-grown LMO (FSLMO with $K = 0$ mm$^{-1}$ and that with $K = 1$ mm$^{-1}$) for quantitively investigating the lattices, octahedral distortion, and octahedral tilting along pseudo-tetragonal [100] and [001] direction. The inset in Figure 3i (Figure 3j and Figure S7b, Supporting Information) shows the representative supercell of LMO and the MnO$_6$ octahedra in the supercell are presented in Figure 3k (Figure 3l and Figure S7c, Supporting Information). Two adjacent inequivalent MnO$_6$ octahedra (denoted as OC1 and OC2, containing Mn1 and Mn2 atoms, respectively) alternate three-dimensionally in LMO before and after releasing and bending when the OC1 and OC2 is stretched and compressed, respectively, in out-of-plane direction. The difference between OC1 and OC2 octahedra displayed vast evolution after releasing and bending. This difference is, for simplicity, depicted by the ratio of the out-of-plane O-O length $Lz$ in OC1 ($Lz1$) and that in OC2 ($Lz2$). The $Lz1$ and $Lz2$ of as-grown LMO, FSLMO with $K = 0$ mm$^{-1}$, and that with $K = 1$ mm$^{-1}$, evaluated statistically from 20 atomic rows and 40 atomic columns (Figure 3i,j, and Figure S7b, Supporting Information), are shown in Figure 3m-o, respectively. Evidently, the as-grown LMO film showed a minor difference between $Lz1$ and $Lz2$ with average $Lz1/Lz2$ being 1.012 (Figure 3m) and thus presented slight characteristics of P2$_1$/n symmetry. After releasing, however, the FSLMO with $K = 0$ mm$^{-1}$ exhibited apparent difference between $Lz1$ and $Lz2$ with average $Lz1/Lz2$ being 1.077 (Figure 3n) and demonstrated distinct characteristics of P2$_1$/n symmetry. Moreover, the FSLMO with $K = 1$ mm$^{-1}$ presented even larger average $Lz1/Lz2$ (1.113, Figure 3o) than that of FSLMO with $K = 0$ mm$^{-1}$ which revealed that the characteristics of P2$_1$/n symmetry of FSLMO got further enhanced from bending. This enhancement in the characteristics of P2$_1$/n symmetry originated from the out-of-plane lattice shrinking and the shortening of the distance between adjacent Mn1 and Mn2 atoms ($L_{\text{Mn1-Mn2}}$) along $c$ direction after releasing and bending (Figure 3p). We think that this out-of-plane lattice shrinking after releasing and bending came from the expansion of the in-plane lattices of LMO. Above results were consistent with our first-principles calculations performed on the freestanding LMO with P2$_1$/n symmetry, in which the magnetic moment per Mn in P2$_1$/n LMO increased (Figure 3q) and the distance between the adjacent Mn1 and Mn2 atoms decreased (Figure 3r) monotonously with the out-of-plane shrinking.

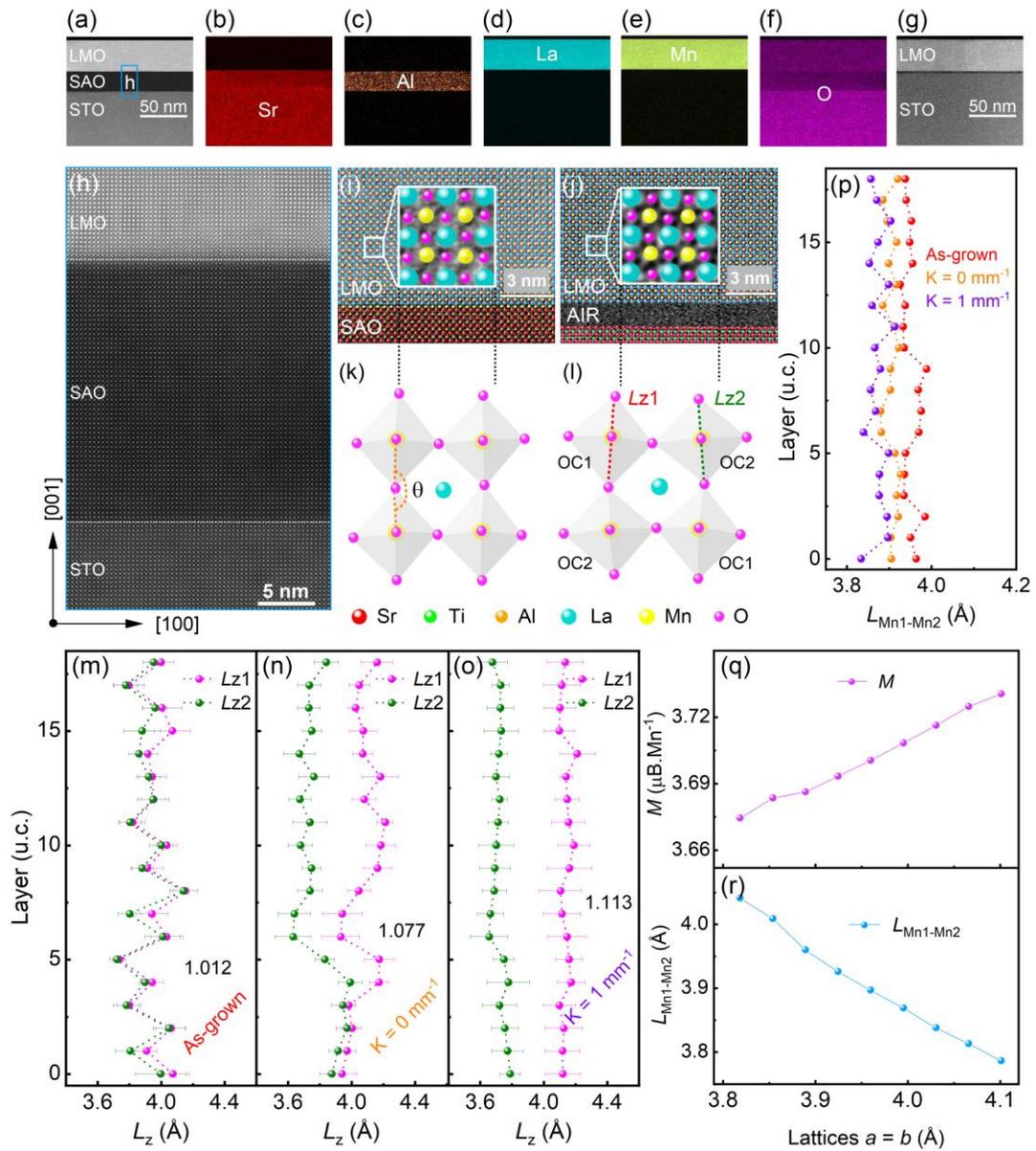

**Figure 3**. a) HAADF cross section of as-grown LMO. b-f) EELS mapping of Sr $L$, Al $K$, La $L$, Mn $K$, and O $K$ edge, respectively, corresponding to (a). g) HAADF cross section of FSLMO with $K$ = 0 mm$^{-1}$. h) Zooming-in of the marked area in (a). The upper and lower dotted lines indicate the LMO-SAO and SAO-STO interfaces, respectively. i, j) Inverted ABF image of as-grown LMO and FSLMO with $K$ = 0 mm$^{-1}$ approaching the LMO-SAO and LMO-AIR interface, respectively. Insets show the magnification of the marked areas in (i) and (j), respectively. k, l) MnO$_6$ octahedra in Insets in (i) and (j), respectively. The red, orange, cyan, yellow and magenta sphere denote Ti, Sr, Al, La, Mn, and O atom, respectively. m-o) $Lz1$ and $Lz2$ of the as-grown LMO, FSLMO with $K$ = 0 mm$^{-1}$, and that with $K$ = 1 mm$^{-1}$, respectively. The numbers denote the average $Lz1/Lz2$. p) The $L_{Mn1-Mn2}$ along $c$ direction in the as-grown LMO, FSLMO with $K$ = 0 mm$^{-1}$, and that with $K$ = 1 mm$^{-1}$. q, r) Magnetic moment per Mn and the $L_{Mn1-Mn2}$ with in-plane strain from first-principles calculations.

The underlying mechanism between the enhancement of the characteristic of P2$_1$/n symmetry and the enhancement of the ferromagnetism in LMO is schematically illustrated in **Figure 4**. Since the ferromagnetic interaction in $ab$ plane is irrespective of the magnitude of distortion,[29,32] we focus on the ferromagnetic interaction along $c$ direction. In the P2$_1$/n structure, the single $e_g$ electron of Mn1 (Mn2) occupies the $d_{3z^2-r^2}$ ($d_{x^2-y^2}$) orbital for lower energy.[12,13] The augmentation of the characteristics of P2$_1$/n symmetry brings stronger out-of-plane stretching (compression) on OC1 (OC2) (Figure 4a,d) and thus produces stronger feature of $d_{3z^2-r^2}/d_{x^2-y^2}$ orbital alternation. Meanwhile, this augmentation of the P2$_1$/n feature draws closer adjacent Mn1 and Mn2 atoms along $c$ direction (Figure 4b,e), which brings larger overlaps between the Mn1 half-filled $d_{3z^2-r^2}$ and the middle O $p_z$ orbitals and larger overlap between the middle O $p_z$ and Mn2 empty $d_{3z^2-r^2}$ orbitals in the FSLMO, comparing to those in the as-grown LMO (Figure 4c,f), so that it boosts ferromagnetism in FSLMO according to the Goodenough-Kanamori rules.[12,33-36]

Although the bended FSLMO membranes were supposed to be compressed at the top and stretched at the bottom in out-of-plane direction when transferred onto quartz rod with nonzero curvature,[7,37] within appropriate curvature range, as in this case, the present FSLMO films were still able to exhibit overall out-of-plane lattice shrinking with increasing flexure degree, which was suggested by the monotonous decrease in $c$ from XRD analyses (Figure 1d) and the softening of B$_{3g}$ stretching mode in Raman spectra (Figure 2f)[26]. This lattice shrinking brings further augmented characteristics of P2$_1$/n structure comparing to the releasing process and leads to monotonous enhancement of $d_{3z^2-r^2}/d_{x^2-y^2}$ orbital alternation and continuous decrease of Mn1-Mn2 distance along $c$ direction, eventually resulting in stronger ferromagnetism with increasing curvature. Such an enhancement originating from the augmented Mn-Mn interaction is similar to the theoretical prediction for some other materials on the flexomagnetism enhancement.[38,39] Additionally, the octahedra of LMO we observed from STEM showed a magnified tilting, which was described by the intersection angle between adjacent octahedra ($\theta_{Mn1-O-Mn2}$) along $c$ direction, after releasing and bending (Figure S8a, Supporting Information). This heavier tilting was responsible for the larger resistivity of LMO after releasing and bending,[29] as shown in Figure S8b, Supporting Information.

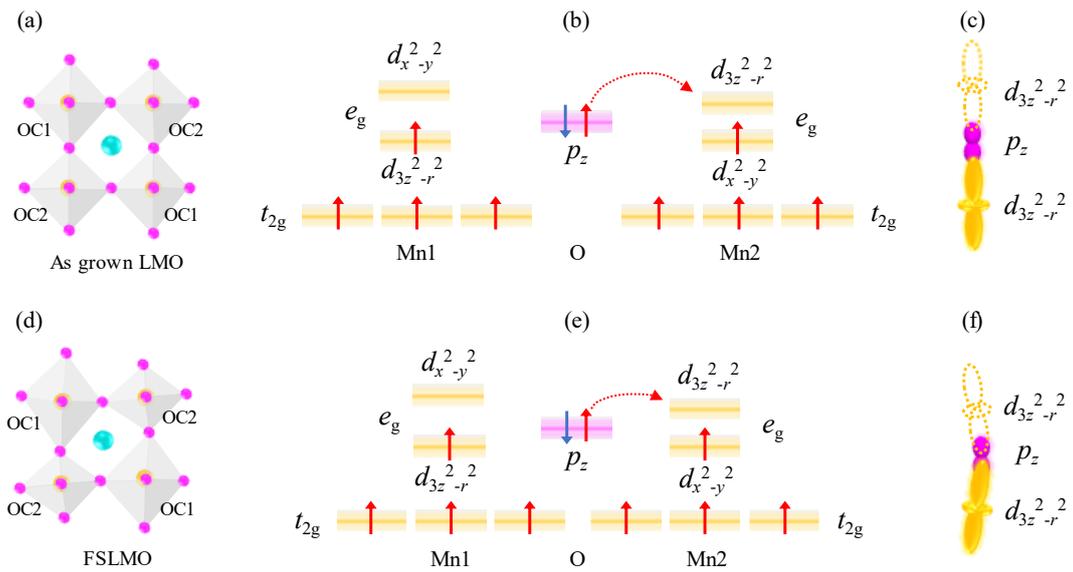

**Figure 4**. a) Schematic illustration of the octahedra in as-grown LMO. b) Occupation of electrons in $d$ orbital of Mn1 and Mn2 atoms and in $p_z$ orbital of middle O atom in as-grown LMO. c) Illustration of the overlaps among the Mn1 half-filled $d_{3z^2-r^2}$, the middle O $p_z$, and the Mn2 empty $d_{3z^2-r^2}$ orbitals in as-grown LMO. d-f) Corresponding octahedral illustration, electron state, and orbital overlaps in FSLMO, respectively.

## 3. Conclusion

In summary, we have realized a remarkable enhancement of the saturation magnetization for LMO films by 56% and 92%, respectively, after releasing LMO from substrates and after bending films to a curvature of 1 mm$^{-1}$ using our water-assisted direct-transferring method. We attribute that the enhanced ferromagnetism is originated from the augmented characteristics of the unconventional P2$_1$/n structure in LMO after the strain release and bending due to the out-of-plane lattice shrinking which is supported by STEM observations and the first-principles calculations. The augmented characteristics of the P2$_1$/n structure brings larger out-of-plane stretching (compression) on OC1 (OC2), stronger feature of $d_{3z^2-r^2}/d_{x^2-y^2}$ orbital alternation, and shorter Mn1-Mn2 distance along $c$ direction, which produces larger overlaps among the Mn1 half-filled $d_{3z^2-r^2}$, the middle O $p_z$, and the Mn2 empty $d_{3z^2-r^2}$ orbitals, and results in a significant enhancement of the ferromagnetism. Our work paves a pathway to effectively manipulate and improve the functionalities of oxides and offer their promising applications in twistronics, flexoelectronics, and wearable devices.

## 4. Experimental Section

*Sample Synthesis*: The STO substrates were etched in 4% HF solution for 40 seconds and were then annealed at 1050 °C for 90 minutes to achieve clear steps and terraces. During the growth with pulsed laser deposition (PLD), a XeCl laser with a central wavelength of 308 nm was employed. The water-sacrificial SAO layer was grown onto the TiO$_2$-terminated STO at 750 °C under an oxygen partial pressure of 5 Pa with a substrate-target distance of 7 cm. The LMO layer was subsequently deposited at same temperature under an optimized oxygen pressure of 0.5 Pa and a substrate-target distance of 8 cm. The laser repetition frequency and energy fluence for SAO were 5 Hz and 2 J/cm$^2$, respectively, and were 3 Hz and 1.5 J/cm$^2$, respectively, for LMO. After the growth, the heterostructure was annealed *in-situ* at the grown conditions of LMO layer for 5 minutes to maintain surface stoichiometry and then was cooled down to room temperature at a rate of 25 °C/min.

*Sample Processing*: Prior to releasing, the 10*10 mm$^2$ LMO/SAO/STO heterostructure was cut into four pieces to prepare for the as-grown LMO, FSLMO with $K = 0$ mm$^{-1}$, that with 0.5 mm$^{-1}$, and that with 1 mm$^{-1}$ sample, respectively, to eliminate the sample dependence in property characterizations. The crack-and-wrinkle-free transferred freestanding films were shaped into rectangle parallel to quartz rods carefully under optical microscope to evade the error in sample size determination. The residual fragments around the sample edge from shaping were completely removed using dedicated wipes with alternative ethanol and acetone.

*X-ray Characterizations*: The XRD and RSM analyses were conducted at room temperature using an ultra-high resolution (PB-Ge (220) × 4) SmartLab Rigaku X-ray diffractometer with the central wavelength of the X-ray (Cu Kα) was 1.5409 Å.

*Magnetic Properties*: The MH loops of all LMO samples were measured at 10 K using a vibrating sample magnetometer (VSM) with physical property measurement system (PPMS) within magnetic field from -15000 to +15000 Oe. The corresponding MT were measured at a rate of 6 K/min under 1000 Oe. The magnetometer was operated at a vibrating frequency of 40 Hz and a vibrating amplitude of 2 mm, which corresponded to a moment measuring range up to 0.01 emu. To the 25 and 50 u.c. sample in Figure S5, the MH loop and MT of the as-grown samples were firstly measured. After that, those samples were immersed into deionized water with sample stick for full etching of SAO buffer layer. After removing SAO utterly, these samples were dried and put into PPMS again for MH and MT measurements.

*Electrical Properties*: For $\rho$ detection, four electrodes made of indium were manufactured at the square corner of the sample used for above magnetic measurements. Each sample was measured under van der Pauw configuration using PPMS from 390 K to 10 K at a set current of 1 μA.

*Raman Spectra*: The Raman spectra were collected at room temperature using a WiTec system. The incident laser beam was focused into a spot with radius of ~1 um on the sample using an optical focusing system. The wavelength of incident beam was

532 nm.

*Scanning Transmission Electron Microscopy*: The atomic structures of the as-grown LMO, FSLMO with $K = 0$ mm$^{-1}$, and that with $K = 1$ mm$^{-1}$ were characterized at room temperature using an ARM-200CF transmission electron microscope which was operated at 200 keV and was equipped with double spherical aberration (Cs) correctors. The LMO layer of the FSLMO with $K = 0$ mm$^{-1}$ was transferred onto a (001) STO substrate which can be taken as a reference when determine the bonding length of LMO. For the FSLMO with $K = 1$ mm$^{-1}$, another (001) STO substrate was prepared simultaneously as a reference for the determination of bonding length. Those samples were firstly covered with successive layer of C and Pt before being mechanically polished to ~20 μm, and were further optimized by precise argon-ion milling to transparent to electron beam. All STEM images used were filtered using the HREM-Filters Pro/Lite developed by HREM Research Inc.

*Electron Energy Loss Spectroscopy*: The unit-cell-resolved EELS spectra of the as-grown LMO, FSLMO with $K = 0$ mm$^{-1}$, and that with $K = 1$ mm$^{-1}$ were collected at room temperature along with the atomic structural characterization using STEM. The EELS data was extracted using DigitalMicrograph software from Gatan Inc.

*Statistical Analyses*: The coordinate of each atom in STEM images in Figure 3i, 3j, and S7b was obtained using the CalAtom[40,41] and the digitizer in OriginPro from OriginLab corporation. Few coordinates acquired automatically which deviate from the atomic center had been manually corrected. Using the coordinates, we evaluated the lattice parameters, $Lz$, $L_{Mn1\text{-}Mn2}$, and $\theta_{Mn1\text{-}O\text{-}Mn2}$.

*First-principles Calculations*: The calculations were performed using density functional theory (DFT) with Vienna *ab initio* Simulation Package (VASP).[42,43] Two alternative LaO and MnO$_2$ layers, combined with top and underneath vacuum layer, were constructed for the simulation of freestanding P2$_1$/n LMO. The vacuum length was taken as 15 Å to avoid the interaction between the periodic slab. The generalized gradient approximation (GGA) exchange-correlation interaction within the Perdew-Burke-Ernzerhof (PBE) was adopted in these DFT calculations.[44] The Projector Augmented Wave (PAW) method was used[45] with the following electronic configurations: $5d^1 6s^2$ (La), $3d^6 4s^1$ (Mn), and $2s^2 2p^4$ (O). A 520 eV energy cutoff of the plane-wave basis set was used for all calculations. For structural optimizations, atomic positions were relaxed until the force on each atom was less than 0.05 eV / Å. A 7 × 7 × 1 Γ-centered k-point mesh was used for structural relaxation, and a denser 20 × 20 × 1 K-mesh was used for electronic structure calculations. We used GGA + U method[46] ($U$ = 3.33 eV and $J$ = 1.33 eV for Mn) to account for on-site Coulomb interactions according to previous theoretical research.[47] The atomic structure was visualized using the Visualization for Electron and Structural Analysis (VESTA) package.[48]

**Supporting Information**

Supporting Information is available from the Wiley Online Library or from the author.

**Acknowledgment**

This work was supported by the National Key Basic Research Program of China (Grant Nos. 2019YFA0308500 and 2020YFA0309100), the National Natural Science Foundation of China (Grant Nos. 11721404, 51761145104, and 11974390), the Youth Innovation Promotion Association of the Chinese Academy of Sciences (Grant No. 2018008), the Beijing Nova Program of Science and Technology (Grant No. Z191100001119112), the Beijing Natural Science Foundation (Grant No. 2202060), and the Strategic Priority Research Program (B) of the Chinese Academy of Sciences (Grant No. XDB33030200).

**Conflict of Interest**

The authors declare no conflict of interest.

**References**

[1] Y. Cao, V. Fatemi, S. Fang, K. Watanabe, T. Taniguchi, E. Kaxiras, P. Jarillo-Herrero. *Nature* **2018**, 556, 43.

[2] Y. Cao, D. Rodan-Legrain, O. Rubies-Bigorda, J. M. Park, K. Watanabe, T.

Taniguchi, P. Jarillo-Herrero. *Nature* **2020**, 583, 215.

[3] K. Yasuda, X. Wang, K. Watanabe, T. Taniguchi, P. Jarillo-Herrero. *Science* **2021**, 372, 1458.

[4] L. L. Shu, S. M. Ke, L. F. Fei, W. B. Huang, Z. G. Wang, J. H. Gong, X. N. Jiang, L. Wang, F. Li, S. J. Lei, Z. G. Rao, Y. B. Zhou, R. K. Zheng, X. Yao, Y. Wang, M. Stengel, G. Catalan. *Nat. Mater.* **2020**, 19, 605.

[5] P. Gao, S. Z. Yang, R. Ishikawa, N. Li, B. Feng, A. Kumamoto, N. Shibata, P. Yu, Y. Ikuhara. *Phys. Rev. Lett.* **2018**, 120, 267601.

[6] D. Lu, D. J. Baek, S. S. Hong, L. F. Kourkoutis, Y. Hikita, H. Y. Hwang. *Nat. Mater.* **2016**, 15, 1255.

[7] G. Dong, S. Li, M. Yao, Z. Zhou, Y.-Q. Zhang, X. Han, Z. Luo, J. Yao, B. Peng, Z. Hu, H. Huang, T. Jia, J. Li, W. Ren, Z.-G. Ye, X. Ding, J. Sun, C.-W. Nan, L.-Q. Chen, J. Li, M. Liu. *Science* **2019**, 366, 475.

[8] F. An, K. Qu, G. Zhong, Y. Dong, W. Ming, M. Zi, Z. Liu, Y. Wang, B. Qi, Z. Ding, J. Xu, Z. Luo, X. Gao, S. Xie, P. Gao, J. Li. *Adv. Func. Mater.* **2020**, 30, 2003495.

[9] H. Elangovan, M. Barzilay, S. Seremi, N. Cohen, Y. Jiang, L. W. Martin, Y. Ivry. *ACS Nano* **2020**, 14, 5053.

[10] R. Xu, J. Huang, E. S. Barnard, S. S. Hong, P. Singh, E. K. Wong, T. Jansen, V. Harbola, J. Xiao, B. Y. Wang, S. Crossley, D. Lu, S. Liu, H. Y. Hwang. *Nat. Commun.* **2020**, 11, 3141.

[11] Q. Wang, H. Fang, D. Wang, J. Wang, N. Zhang, B. He, W. Lü. *Crystals* **2020**, 10, 733.

[12] Y. S. Hou, H. J. Xiang, X. G. Gong. *Phys. Rev. B* **2014**, 89, 064415.

[13] M. Yang, K. Jin, H. Yao, Q. Zhang, Y. Ji, L. Gu, W. Ren, J. Zhao, J. Wang, E. J. Guo, C. Ge, C. Wang, X. Xu, Q. Wu, G. Yang. *Adv. Sci.* **2021**, 8, 2100177.

[14] Y. K. Liu, H. F. Wong, K. K. Lam, C. L. Mak, C. W. Leung. *J. Magn. Magn. Mater.* **2019**, 481, 85.

[15] Y. M. Liang, Z. J. Wang, Y. Bai, Y. J. Wu, X. K. Ning, X. G. Zhao, W. Liu, Z. D. Zhang. *J. Phys. Chem. C* **2019**, 123, 14842.

[16] R. K. Zheng, Y. Wang, H. U. Habermeier, H. L. W. Chan, X. M. Li, H. S. Luo. *J. Alloys Compd.* **2012**, 519, 77.

[17] A. M. Zhang, S. L. Cheng, J. G. Lin, X. S. Wu. *J. Appl. Phys.* **2015**, 117, 17b325.

[18] R. K. Zheng, H. U. Habermeier, H. L. W. Chan, C. L. Choy, H. S. Luo. *Phys. Rev. B* **2010**, 81, 104427

[19] D. J. Baek, D. Lu, Y. Hikita, H. Y. Hwang, L. F. Kourkoutis. *ACS Appl. Mater. Inter.* **2017**, 9, 54.

[20] Z. Lu, J. Liu, J. Feng, X. Zheng, L.-h. Yang, C. Ge, K.-j. Jin, Z. Wang, R.-W. Li. *APL Mater.* **2020**, 8, 051105.

[21] D. Ji, S. Cai, T. R. Paudel, H. Sun, C. Zhang, L. Han, Y. Wei, Y. Zang, M. Gu, Y. Zhang, W. Gao, H. Huyan, W. Guo, D. Wu, Z. Gu, E. Y. Tsymbal, P. Wang, Y. Nie, X. Pan. *Nature* **2019**, 570, 87.

[22] H. Y. Sun, C. C. Zhang, J. M. Song, J. H. Gu, T. W. Zhang, Y. P. Zang, Y. F. Li, Z. B. Gu, P. Wang, Y. F. Nie. *Thin Solid Films* **2020**, 697, 137815.

[23] M. Li, C. Tang, T. R. Paudel, D. Song, W. Lü, K. Han, Z. Huang, S. Zeng, X. R. Wang, A. Ping Yang, J. Chen, T. Venkatesan, E. Y. Tsymbal, C. Li, S. J. Pennycook. *Adv. Mater.* **2019**, 31, 1901386.

[24] Z. Chen, Z. Chen, Z. Q. Liu, M. E. Holtz, C. J. Li, X. R. Wang, W. M. Lu, M. Motapothula, L. S. Fan, J. A. Turcaud, L. R. Dedon, C. Frederick, R. J. Xu, R. Gao, A. T. N'Diaye, E. Arenholz, J. A. Mundy, T. Venkatesan, D. A. Muller, L. W. Wang, J. Liu, L. W. Martin. *Phys. Rev. Lett.* **2017**, 119, 156801.

[25] T. C. Kaspar, P. V. Sushko, S. R. Spurgeon, M. E. Bowden, D. J. Keavney, R. B. Comes, S. Saremi, L. Martin, S. A. Chambers. *Adv. Mater. Interfaces* **2018**, 6, 1801428.

[26] A. Chaturvedi, V. Sathe. *Thin Solid Films* **2013**, 548, 75.

[27] M. N. Iliev, M. V. Abrashev, H.-G. Lee, V. N. Popov, Y. Y. Sun, C. Thomsen, R. L. Meng, C. W. Chu. *Phys. Rev. B* **1998**, 57, 2872.

[28] E. Granado, J. A. Sanjurjo, C. Rettori, J. J. Neumeier, S. B. Oseroff. *Phys. Rev. B* **2000**, 62, 11304.

[29] C. Zhong, X. Lu, Y. Wan, Y. Min, Z. Zhao, P. Zhou, Z. Dong, J. Liu. *J. Magn. Magn. Mater.* **2018**, 466, 406.

[30] B. R. K. Nanda, S. Satpathy. *J. Magn. Magn. Mater.* **2010**, 322, 3653.

[31] Z. L. Wang, J. S. Yin, Y. D. Jiang, J. Zhang. *Appl. Phys. Lett.* **1997**, 70, 3362.

[32] I. Solovyev, N. Hamada, K. Terakura. *Phys. Rev. Lett.* **1996**, 76, 4825.

[33] J. S. Zhou, J. B. Goodenough. *Phys. Rev. Lett.* **2006**, 96, 247202.

[34] P. W. Anderson. *Phys. Rev.* **1950**, 79, 350.

[35] J. B. Goodenough. *Phys. Rev.* **1955**, 100, 564.

[36] J. B. Goodenough, A. L. Loeb. *Phys. Rev.* **1955**, 98, 391.

[37] G. Dong, S. Li, T. Li, H. Wu, T. Nan, X. Wang, H. Liu, Y. Cheng, Y. Zhou, W. Qu, Y. Zhao, B. Peng, Z. Wang, Z. Hu, Z. Luo, W. Ren, S. J. Pennycook, J. Li, J. Sun, Z. G. Ye, Z. Jiang, Z. Zhou, X. Ding, T. Min, M. Liu. *Adv. Mater.* **2020**, 32, 2004477.

[38] Y.-H. Shen, Y.-X. Song, W.-Y. Tong, X.-W. Shen, S.-j. Gong, C.-G. Duan. *Adv. Theory Simul.* **2018**, 1, 1800048.

[39] P. Lukashev, R. F. Sabirianov. *Phys. Rev. B* **2010**, 82, 094417.

[40] Q. Zhang, L. Y. Zhang, C. H. Jin, Y. M. Wang, F. Lina. *Ultramicroscopy* **2019**, 202, 114.

[41] Q. Zhang, C. H. Jin, H. T. Xu, L. Y. Zhang, X. B. Ren, Y. Ouyang, X. J. Wanga, X. J. Yue, F. Lin. *Micron* **2018**, 113, 99.

[42] G. Kresse, J. Furthmuller. *Phys. Rev. B* **1996**, 54, 11169.

[43] G. Kresse, J. Furthmuller. *Comput. Mater. Sci.* **1996**, 6, 15.

[44] J. P. Perdew, A. Ruzsinszky, G. I. Csonka, O. A. Vydrov, G. E. Scuseria, L. A. Constantin, X. Zhou, K. Burke. *Phys. Rev. Lett.* **2008**, 100, 136406.

[45] P. E. Blochl. *Phys. Rev. B* **1994**, 50, 17953.

[46] A. I. Liechtenstein, V. V. Anisimov, J. Zaanen. *Phys. Rev. B* **1995**, 52, R5467.

[47] M. An, Y. Weng, H. Zhang, J. Zhang, Y. Zhang, S. Dong. *Phys. Rev. B* **2017**, 96, 235112.

[48] K. Momma, F. Izumi. *J. Appl. Cryst.* **2011**, 44, 1272.